%% file: siggraphconferencepapers26-100.tex
\documentclass[acmtog]{acmart}

\usepackage{booktabs} 

\copyrightyear{2026}
\acmYear{2026}
\setcopyright{cc}
\setcctype{by}
\acmConference[SIGGRAPH Conference Papers '26]{Special Interest Group on Computer Graphics and Interactive Techniques Conference Conference Papers}{July 19--23, 2026}{Los Angeles, CA, USA}
\acmBooktitle{Special Interest Group on Computer Graphics and Interactive Techniques Conference Conference Papers (SIGGRAPH Conference Papers '26), July 19--23, 2026, Los Angeles, CA, USA}
\acmDOI{10.1145/3799902.3811137}
\acmISBN{979-8-4007-2554-8/2026/07}

\begin{document}
\title{Non-line-of-sight imaging with arbitrary relay surface geometries via 3D Gaussian Transient Rendering}
\author{Yi Wang}
\authornote{Both authors contributed equally to this research.}
\orcid{0009-0005-3160-5189}
\email{wangyi24@mails.tsinghua.edu.cn}
\affiliation{%
  \institution{Department of Precise Instrument, Tsinghua University}
  \city{Beijing}
  \country{China}
}
\author{Ziyu Zhan}
\orcid{0009-0004-1951-1043}
\authornotemark[1]
\email{zhanzy21@mails.tsinghua.edu.cn}
\affiliation{%
  \institution{Department of Precise Instrument, Tsinghua University}
  \city{Beijing}
  \country{China}
}

\author{Yuran Wang}
\orcid{0009-0003-4961-6658}
\email{wangyuran21@mails.tsinghua.edu.cn}
\affiliation{%
  \institution{Department of Precise Instrument, Tsinghua University}
  \city{Beijing}
  \country{China}
  }

\author{Hao Wang}
\email{wh18810520679@163.com}
\orcid{0000-0003-0210-3372}
\affiliation{%
  \institution{Department of Electrical Engineering, City University of Hong Kong}
  \city{Hong Kong}
  \country{China}
}

\author{Qiang Liu}
\email{qiangliu@tsinghua.edu.cn}
\orcid{0000-0002-1638-7567}
\affiliation{%
 \institution{Department of Precise Instrument, Tsinghua University}
 \city{Beijing}
 \country{China}}

\author{Zuoqiang Shi}
\email{zqshi@tsinghua.edu.cn}
\orcid{0000-0002-9122-0302}
\affiliation{%
  \institution{Yau Mathematical Science Center, Tsinghua University}
  \city{Beijing}
  \country{China}}

\author{Lingyun Qiu}
\affiliation{%
  \institution{Yau Mathematical Science Center, Tsinghua University}
  \city{Beijing}
  \country{China}}
\email{lyqiu@tsinghua.edu.cn}
\orcid{0000-0002-2204-7235}

\author{Xing Fu}
\authornote{Corresponding author.}
\affiliation{%
  \institution{Department of Precise Instrument, Tsinghua University}
  \city{Beijing}
  \country{China}}
\email{fuxing@tsinghua.edu.cn}
\orcid{0000-0003-1758-1561}

\begin{abstract}
Imaging objects hidden outside the direct line of sight expands the effective field of view and is critical for applications such as autonomous driving and robotic perception. Despite impressive progress in time-of-flight (ToF)-based non-line-of-sight (NLOS) imaging, real-world deployment remains challenging because practical measurements are often collected over spatially limited, arbitrarily shaped relay regions—conditions that violate the planar-wall and dense-sampling assumptions made by most existing methods.
 To address these limitations, we propose a LOS-guided NLOS imaging pipeline that imposes no geometric assumptions on the relay surface and naturally supports both confocal and non-confocal configurations. Our method represents the hidden scene using 3D Gaussian primitives and couples them with an efficient, differentiable transient rendering model, enabling end-to-end optimization directly from measured transients. We validate our approach on real-world measurements from both a public dataset and a custom-built capture system. Across settings, our method achieves state-of-the-art reconstruction fidelity under spatially limited, sparsely sampled conditions, and significantly outperforms existing methods on complex, arbitrary relay surface geometries. We release our code and datasets at \href{https://github.com/YWnlos/nlos-3d-gtr}{nlos-3d-gtr}.
\end{abstract}

\ccsdesc[500]{Computing methodologies~Computational photography}
\ccsdesc[500]{Computing methodologies~Rendering}

\begin{teaserfigure}
  \includegraphics[width=\textwidth]{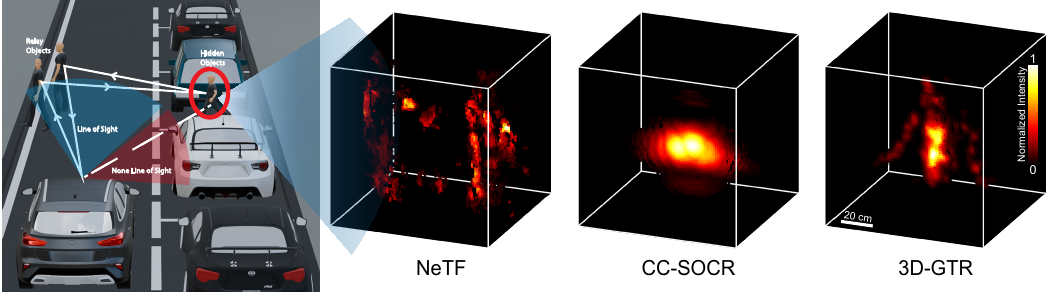}
  \caption{\textbf{Non-line-of-sight (NLOS) imaging with arbitrary relay surface geometries.} 
 (Left) A schematic illustration of a ``pedestrian dart-out'' hazard. In this scenario, \textbf{NLOS imaging} is crucial for detecting the pedestrian emerging from a blind spot to prevent potential accidents. In the absence of a large planar wall, the system utilizes the backs of two visible pedestrians as relay surfaces. These surfaces are non-planar and spatially limited with complex normal distributions, making traditional NLOS imaging methods inapplicable.
  (Right) Comparison of reconstruction results using NeTF~\cite{shen2021non}, CC-SOCR~\cite{liu2023non}, and our proposed 3D
Gaussian Transient Rendering (3D-GTR). While other state-of-the-art methods fail to resolve the target under such complex and irregular geometric constraints, our approach robustly recovers the hidden pedestrian geometry. The left scene illustration contains modified 3D assets from Sketchfab: ``Suv Car'' by appsnation, ``Car'' by pavan.amalakanti, ``Stylized 3D Car Pack'' by DevPoly3D, and ``Car'' by Tech developers, all used under CC BY 4.0.}
  \Description{The figure contains a schematic illustration of a ``pedestrian dart-out'' hazard and the comparison reconstructions among our methods and other state-of-the-art methods.}
  \label{fig:teaser}
\end{teaserfigure}

\maketitle

\input{main}


\end{document}

%% file: main.tex
\section{Introduction}
Imaging objects hidden from a sensor’s direct line of sight (LOS) remains a long-standing challenge. Among various non-line-of-sight (NLOS) imaging scenarios, 
time-of-flight (ToF)-based imaging, which aims to ``look around corners'' by exploiting a visible relay surface while the target is occluded, has attracted significant interest in remote sensing, robotic perception, and public safety.
To meet these demands, a wide range of hardware advances~\cite{buttafava2015non,scheiner2020seeing} and software algorithms~\cite{o2018confocal,liu2019non} have been proposed. Furthermore, because transient ToF acquisition shares a similar system architecture with conventional LiDAR, it offers a promising route toward practical integration into automotive sensing systems~\cite{rapp2020advances}. 
Despite this impressive progress, bringing NLOS imaging to real-world deployment remains difficult. 
Most existing methods assume a large, continuous, and planar relay wall. In practice, however, transient measurements are often acquired from spatially limited regions with arbitrary geometries.
To address these challenges, researchers have explored several directions, including model-based iterative frameworks~\cite{liu2023non}, learning-based approaches~\cite{cui2025transdiff}, and rendering-driven neural-field reconstruction pipelines~\cite{shen2021non}. Nevertheless, these methods face significant practical hurdles: they often require prolonged (hours) training or optimization times and lack robustness in complex environments, especially when the relay surface is unconstrained and non-planar. This highlights the critical need for a fast, reliable NLOS reconstruction framework capable of generalizing across arbitrary relay surface geometries. 
To this end, we propose a LOS-guided NLOS imaging pipeline that imposes no geometric assumptions on the relay surface. By leveraging LOS measurements, we recover not only positions but also the local geometry of the relay surface, which in turn guides the NLOS detection and reconstruction process. Inspired by the success of 3DGS in novel view synthesis (NVS)~\cite{kerbl20233d}, and 3D reconstruction~\cite{guedon2024sugar,keetha2024splatam}, we represent the hidden scene using 3D Gaussian primitives. We couple the Gaussian primitives with a physically grounded, differentiable transient renderer and optimize them via backpropagation to minimize the loss between predicted and measured transient signals. Once optimized, the recovered scene representation can be fed into the transient renderer to synthesize dense, physically consistent transient measurements, which can be subsequently coupled with conventional transient solvers to obtain the final reconstruction. Extensive experiments on real-world measurements—from both public datasets and our custom capture system—demonstrate the robustness of our approach across a wide range of settings, including confocal planar, non-confocal planar, and non-confocal scenarios with arbitrarily shaped relay surfaces. The main contributions of our work are as follows:
\begin{itemize}
    \setlength{\itemsep}{1pt}
      \setlength{\topsep}{2pt}
    \setlength{\parsep}{0pt}
     \setlength{\partopsep}{0pt}
    \item We propose a LOS-guided NLOS imaging pipeline (shown in Fig. ~\ref{fig:pipeline}) that eliminates the need for prior assumptions about relay surface geometry. Consequently, this formulation naturally supports both confocal and non-confocal acquisitions, as well as arbitrary illumination-detection pair configurations for transient measurements, substantially broadening the practical applicability of NLOS imaging.
    \item We develop 3D Gaussian Transient Rendering (\textbf{3D-GTR}), an efficient, fully differentiable, physics-based transient rendering model built on lightweight 3D Gaussian primitives. By explicitly incorporating physics-aware light transport, our renderer delivers reliable performance across diverse acquisition configurations.
    \item We validate our method on real-world measurements from both public datasets and a custom-built capture system involving diverse relay surfaces. Our results demonstrate significantly improved robustness compared to existing methods, particularly under complex relay surface geometries. 
\end{itemize}
\section{Related Work}
\subsection{NLOS Imaging.}
ToF-based NLOS imaging has witnessed rapid progress since the seminal work of Kirmani \emph{et al.}~\cite{kirmani2009looking}. Early reconstruction methods based on filtered back-projection (FBP)~\cite{velten2012recovering,la2018error} established a practical baseline for real-world NLOS imaging. Subsequently, the light-cone transform (LCT)~\cite{o2018confocal} introduced a convolutional formulation under confocal measurements, significantly reducing computational complexity. In parallel, wave-optics approaches, including f-k migration~\cite{lindell2019wave} and RSD~\cite{liu2020phasor,liu2019non}, model hidden-scene light transport as wave propagation, enabling high-fidelity reconstructions. Recent learning-based methods leverage deep priors learned from data to improve reconstruction quality~\cite{chen2020learned,yu2023enhancing}. In addition, a number of works focus on accelerating reconstruction~\cite{li2022fast,wei2025fast} and further enhancing imaging quality~\cite{zhang2025sub}.

\paragraph{NLOS imaging with arbitrary transients.}
While most standard approaches rely on dense, uniform sampling over large planar relay surfaces, research on reconstruction from \emph{arbitrary transients} remains limited.
To reduce acquisition time via \emph{sparse measurements}, model-based optimization methods---ranging from compressed sensing~\cite{ye2021compressed} and probabilistic regularization~\cite{liu2023few} to the recent convolution-approximation strategy DO-NLOS~\cite{miao2025under}---have been proposed, alongside data-driven approaches using deep priors~\cite{li2023deep,wang2023non} or temporal consistency~\cite{li2024toward,ye2024plug,ye2024real}.
For \emph{irregular spatial sampling patterns}, CC-SOCR~\cite{liu2023non} introduces a ``virtual confocal'' proxy to bridge arbitrary patterns with confocal reconstruction. Virtual Scanning~\cite{cui2024virtual} adopts a self-supervised strategy for irregularly undersampled data, while TransDiff~\cite{cui2025transdiff} leverages latent diffusion models to ``hallucinate'' dense measurements from aperture-limited inputs; extreme sparsity is explored in single-path keyhole imaging~\cite{metzler2020keyhole}.
To handle \emph{non-planar relay geometries}, 3D RSD~\cite{gu2023fast} extends wave-based solvers via two-stage propagation with a planar proxy, whereas other solutions rely on hardware-aided surface tracking~\cite{la2020non} or accelerated FBP variants~\cite{sun2026cuda}.

\begin{figure*}[t]
  \centering
  \includegraphics[width=0.75\linewidth]{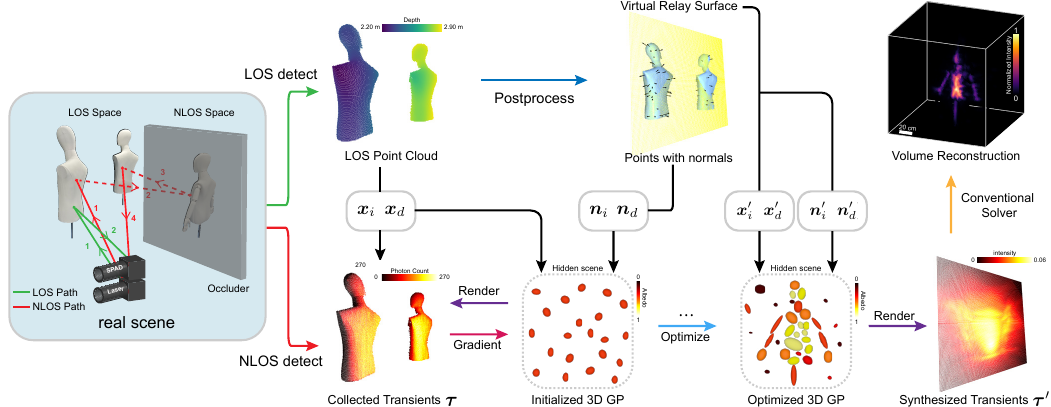}
  \caption{\textbf{Overview of our LOS-guided NLOS imaging pipeline.} From the real scene, we acquire LOS ToF measurements (green path) to obtain a relay surface point cloud (colored by depth), and then perform NLOS detection (red path) at these known locations $\boldsymbol{x}_i, \boldsymbol{x}_d$ to obtain collected transients $\boldsymbol{\tau}$ (colored by photon counts). We then postprocess the LOS point cloud to estimate surface normals and generate a virtual relay surface (yellow). The hidden scene is represented by a set of 3D Gaussian primitives (3D GP). By leveraging our differentiable rendering model with the measured geometry ($\boldsymbol{x}_{i,d}, \boldsymbol{n}_{i,d}$), we iteratively optimize the Gaussian parameters (position, scale, orientation, and albedo) to match the collected transients. The optimized 3D GP are then used to render synthesized transients $\boldsymbol{\tau}'$ (colored by signal intensity) on the virtual plane ($\boldsymbol{x}'_{i,d}, \boldsymbol{n}'_{i,d}$), enabling final volume reconstruction via conventional solvers.}
  \label{fig:pipeline} 
  \Description{The figure contains the LOS-guided NLOS imaging pipeline.}
\end{figure*}

\paragraph{NLOS imaging via Transient Rendering.}
``Analysis-by-synthesis'' frameworks reconstruct the scene by optimizing a physical representation to minimize the error between simulated and measured transients. With solid theoretical foundations established for transient light transport~\cite{jarabo2014framework,pediredla2019ellipsoidal,wu2021differentiable,yi2021differentiable}, early works employed explicit mesh representations optimized via backpropagation, evolving from numerical differentiation~\cite{tsai2019beyond,iseringhausen2020non} to analytical derivatives~\cite{plack2023fast,choi2023self}. Beyond explicit surfaces, implicit representations were pioneered by Neural Transient Fields (NeTF)~\cite{shen2021non}, which model the hidden scene as a continuous volumetric density field optimized via ellipsoid integration. This paradigm further incorporates Signed Distance Functions for precise surface extraction~\cite{grau2022occlusion,fujimura2023nlos}, and has further been integrated with holistic priors~\cite{huang2023omni,shen2024holi} or physics-guided encoders~\cite{mu2022physics} to enhance generalization, while recent efforts employ domain pruning strategies to accelerate convergence~\cite{shim2024domain}.

\subsection{3D Gaussian Splatting and NVS.}
Synthesizing transients under arbitrary configurations mirrors NVS. Following the evolution from NeRF~\cite{mildenhall2021nerf} to accelerated volumetric representations~\cite{fridovich2022plenoxels,muller2022instant,lassner2021pulsar}, 3D Gaussian Splatting (3DGS)~\cite{kerbl20233d} has emerged as a powerful paradigm.
Recent works have also explored differentiable 3D representations for LOS transient or active sensing, including LiDAR view synthesis~\cite{malik2023transient}, propagating-light inverse rendering~\cite{malik2025neural}, diffuse LiDAR-RGB scanning~\cite{behari2025blurred}, low-cost single-photon imaging~\cite{mu2024towards}, and few-pixel ToF reconstruction~\cite{sifferman2025recovering}. Gaussian splatting has also been adapted to related sonar and radar imaging models, including camera-sonar fusion via depth-axis splatting~\cite{qu2024z}, imaging-sonar view synthesis~\cite{sethuraman2025sonarsplat}, and radar data synthesis and 3D reconstruction~\cite{kung2025radarsplat}. Together, these works indicate that Gaussian splatting can be effective beyond standard perspective rendering when its splatting process is adapted to the sensing physics. Motivated by this observation, we introduce 3D Gaussian primitives to NLOS transient imaging. This representation is well suited to arbitrary transients because it combines the flexibility of continuous fields with the efficiency of discrete points.

\section{Methods}
\begin{figure}[htbp]
  \centering
  \includegraphics[width=0.75\linewidth]{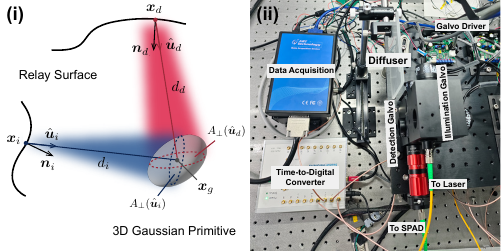}
  \caption{\textbf{Schematic diagram of 3D Gaussian Transient Rendering \& Custom-built experimental setup.} (i) The schematic graph of 3D Gaussian Transient Rendering. (ii) The custom-built experimental setup.}
  \Description{The figure contains schematic graph of 3D Gaussian Transient Rendering and experiment setup.}
  \label{fig:method}
\end{figure}

We consider a standard three-bounce NLOS ToF measurement scenario that places no restrictions on the relay surface geometry and supports both confocal and non-confocal acquisition. An ultrashort laser pulse illuminates a point $\boldsymbol{x}_i$ on the relay surface, and the scattered light propagates into the hidden volume. After interacting with the hidden scene, light returns to the relay surface, where a single-photon avalanche diode (SPAD) records a time-resolved photon-count transient at a detection point $\boldsymbol{x}_d$. As depicted in Fig. \ref{fig:method}, we index each transient with an illumination-detection pair and their local outward normals, i.e.,
$\boldsymbol{\tau}(\boldsymbol{x}_i,\boldsymbol{n}_i,\boldsymbol{x}_d,\boldsymbol{n}_d)$.
For conciseness, we denote the transient associated with the acquisition pair $p$ and its normal $n_p$ as $\boldsymbol{\tau}_{p,n_p}$ in the following.
\subsection{3D Gaussian-Primitive Transient Rendering}
\label{sec:gs_transient_rendering}
We represent the hidden scene using 3D Gaussian primitives, parameterized as
$
\mathcal{G}(\boldsymbol{x}_g,\boldsymbol{\Sigma},\rho)\ \triangleq\ \rho\,\boldsymbol{X}$, $ 
\boldsymbol{X}\sim \mathcal{N}(\boldsymbol{x}_g,\boldsymbol{\Sigma}),
$
where $\boldsymbol{x}_g$ and $\boldsymbol{\Sigma}$ denote the mean and covariance of the spatial distribution, and $\rho$ controls the overall scattering strength.
Physically, each primitive can be interpreted as a spatially localized \emph{scattering density}: a continuum of infinitesimal scatterers concentrated around $\boldsymbol{x}_g$, distributed according to $\boldsymbol{\Sigma}$, and weighted by $\rho$.
Given an illumination-detection pair and the corresponding NLOS transport geometry, our transient rendering model maps each 3D Gaussian primitive to an induced \emph{1D temporal Gaussian kernel} parameterized by $(\mu_t,\sigma_t,A)$, which respectively encode the mean ToF, temporal spread, and radiometric amplitude of its contribution.

\begin{equation}
\tilde{s}_g(t;\boldsymbol{x}_i,\boldsymbol{n}_i,\boldsymbol{x}_d,\boldsymbol{n}_d)
\ \dot{\propto}\ 
A\;\mathcal{N}\big(t;\mu_t,\sigma_t^2\big).
\label{eq:primitive_transient_compact}
\end{equation}
For an illumination-detection relay pair $(\boldsymbol{x}_i,\boldsymbol{n}_i)$ and $(\boldsymbol{x}_d,\boldsymbol{n}_d)$, we define
\[
d_i \triangleq \|\boldsymbol{x}_g-\boldsymbol{x}_i\|_2,
d_d \triangleq \|\boldsymbol{x}_g-\boldsymbol{x}_d\|_2,
\hat{\boldsymbol{u}}_i \triangleq \frac{\boldsymbol{x}_g-\boldsymbol{x}_i}{d_i},
\hat{\boldsymbol{u}}_d \triangleq \frac{\boldsymbol{x}_g-\boldsymbol{x}_d}{d_d}.
\]
The optical transport length for a point $\boldsymbol{X}$ in the hidden scene is
\[
L(\boldsymbol{X}) \triangleq \|\boldsymbol{X}-\boldsymbol{x}_i\|_2+\|\boldsymbol{X}-\boldsymbol{x}_d\|_2.
\]
Assuming the primitive’s spatial spread is small compared to the transport distances, we linearize $L(\boldsymbol{x})$ around $\boldsymbol{x}_g$:
\[
L(\boldsymbol{X}) \approx L(\boldsymbol{x}_g) + \nabla L(\boldsymbol{x}_g)^\top(\boldsymbol{X}-\boldsymbol{x}_g),
\]
where
\[
\nabla L(\boldsymbol{x}_g)
=\frac{\boldsymbol{x}_g-\boldsymbol{x}_i}{\|\boldsymbol{x}_g-\boldsymbol{x}_i\|_2}
+\frac{\boldsymbol{x}_g-\boldsymbol{x}_d}{\|\boldsymbol{x}_g-\boldsymbol{x}_d\|_2}
=\hat{\boldsymbol{u}}_i+\hat{\boldsymbol{u}}_d
\ \triangleq\ \boldsymbol{v}.
\]
Since $\boldsymbol{X}\sim\mathcal{N}(\boldsymbol{x}_g,\boldsymbol{\Sigma})$, the induced path length is (approximately) Gaussian:
$
L(\boldsymbol{X})\ \dot{\sim}\ \mathcal{N}\big(\mu_L,\sigma_L^2\big)$,
$\mu_L = L(\boldsymbol{x}_g)=d_i+d_d$,
$
\sigma_L^2 = \boldsymbol{v}^\top \boldsymbol{\Sigma}\,\boldsymbol{v}.
$

Equivalently, the pushforward response:
\[
\tilde{s}_g(L)\ \propto\ \int_{\mathbb{R}^3} p(\boldsymbol{X})\,\delta\big(L-L(\boldsymbol{X})\big)\,d\boldsymbol{X}
\ \equiv\ p_{L(\boldsymbol{X})}(L),
\]
admits the compact approximation:
\begin{equation}
\tilde{s}_g(L)\ \dot{\propto}\ \mathcal{N}\big(L;\mu_L,\sigma_L^2\big),
\label{eq:length_gaussian}
\end{equation}
where $\mu_L$ and $\sigma_L$ correspond to the center and width of the resulting 1D Gaussian transient in the path-length domain, respectively.

For radiometric weighting (i.e., expected photon counts), we adopt a physically motivated model: the amount of light intercepted by a Gaussian primitive from an illumination point, or collected toward a detector, along direction $\boldsymbol{u}$ at distance $d$, is proportional to the corresponding solid angle:
\begin{equation}
\Omega_{m}\ \propto\ \frac{A_\perp(\hat{\boldsymbol{u}})}{d^2},
\label{eq:solid_angle}
\end{equation}
where $A_\perp(\hat{\boldsymbol{u}})$ denotes the primitive's effective projected area perpendicular to $\boldsymbol{u}$, computed as
$A_\perp(\hat{\boldsymbol{u}})\triangleq
\sqrt{|\boldsymbol{\Sigma}|\big(\hat{\boldsymbol{u}}^\top\boldsymbol{\Sigma}^{-1}\hat{\boldsymbol{u}}\big)}$.

The relay surface angular response can be written in a general bidirectional reflectance distribution function (BRDF)-dependent form.
For the illumination and detection relay points, we denote the corresponding angular weights as
\begin{equation}
    \gamma_i = \mathcal{B}_i(\boldsymbol{\omega}^{\mathrm{in}}_i,
    \boldsymbol{\omega}^{\mathrm{out}}_i;\boldsymbol{n}_i),
    \qquad
    \gamma_d = \mathcal{B}_d(\boldsymbol{\omega}^{\mathrm{in}}_d,
    \boldsymbol{\omega}^{\mathrm{out}}_d;\boldsymbol{n}_d),
\label{eq:relay_brdf}
\end{equation}
where $\boldsymbol{\omega}^{\mathrm{in}}$ and $\boldsymbol{\omega}^{\mathrm{out}}$ denote the incident and outgoing directions at the relay surface, respectively, and $\mathcal{B}_i$ and $\mathcal{B}_d$ describe the BRDF-related angular dependence at the illumination and detection points.
In common NLOS scenarios, the angular weights reduce to non-negative cosine factors under diffuse-relay assumption:
\begin{equation}
    \gamma_i =\max(0,\boldsymbol{n}_i^\top \hat{\boldsymbol{u}}_i),\qquad
    \gamma_d = \max(0,\boldsymbol{n}_d^\top \hat{\boldsymbol{u}}_d).
\label{eq:geometry}
\end{equation}

Based on Eq. \eqref{eq:solid_angle}, the primitive radiometric weight is:
\[
A \;=\; g\,\rho\;\gamma_i\gamma_d\;\frac{A_\perp(\hat{\boldsymbol{u}}_i)\,A_\perp(\hat{\boldsymbol{u}}_d)}{d_i^2\,d_d^2},
\]
where the global gain $g$ absorbs configuration-dependent constants (e.g., detection area/efficiency, unit conversions). The overall radiometric weight can be factorized into three physically grounded processes: illumination-to-Gaussian transport, Gaussian scattering, and Gaussian-to-detection transport. Notably, this radiometric expression yields the familiar inverse fourth-power distance dependence that appears in most NLOS forward models. 
Equivalently,  $\mathcal{N}(L;\mu_L,\sigma_L^2)$ in Eq. \eqref{eq:length_gaussian} induces $\mathcal{N}(t;\mu_L/c_0,\sigma_L^2/c_0^2)$ in Eq. \eqref{eq:primitive_transient_compact} under the light-transport condition: $t=L/c_0$.
After neglecting inter-primitive multiple scattering, for a fixed illumination--detection pair, the synthetic transient $\hat{\boldsymbol{\tau}}_p$ is calculated by the summation of the contributions of all Gaussian primitives.

\begin{figure*}[htbp]
  \centering
  \includegraphics[width=0.75\linewidth]{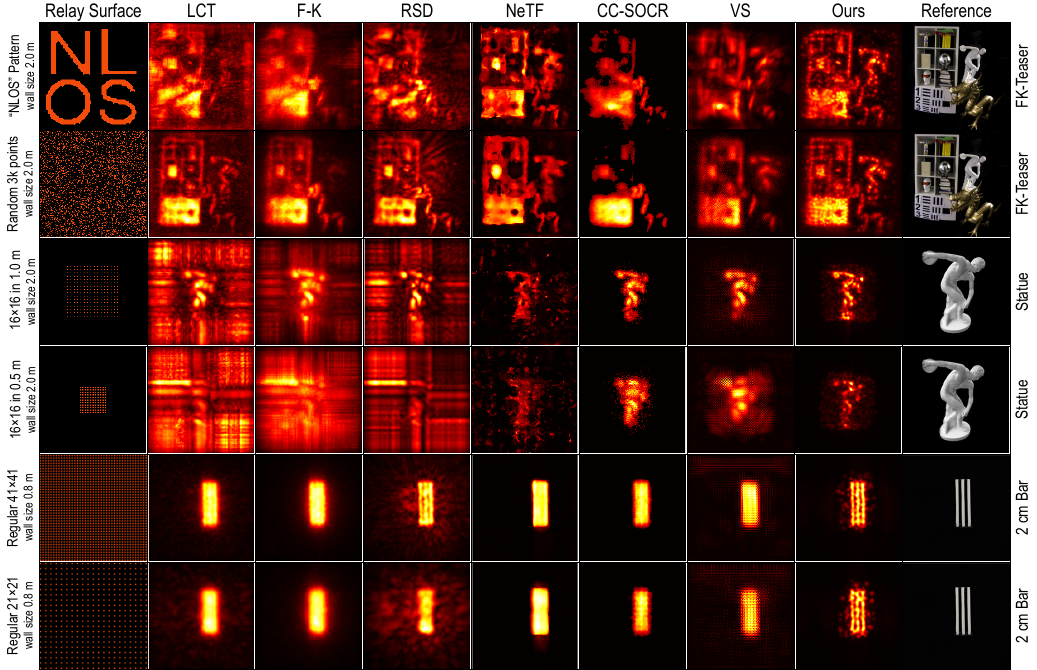}
  \caption{\textbf{Reconstruction results for confocal configuration.} \textbf{Data:} We evaluate three real-world scenes. \textit{Teaser} and \textit{Statue} (from~\cite{lindell2019wave}) were captured on a $2.0\,\text{m} \times 2.0\,\text{m}$ planar relay surface with a dense $128 \times 128$ confocal scan. The \textit{2 cm Bar} scene consists of three $2\,\text{cm}$-wide, $30\,\text{cm}$-long bars ($2\,\text{cm}$ spacing) and was captured by our custom-built system on a $0.8\,\text{m} \times 0.8\,\text{m}$ relay surface sampled on an $81 \times 81$ grid at $16\,\text{ps}$ temporal resolution. Full measurements are masked to obtain region-limited, aperture-limited, or resolution-limited transients. \textbf{Comparison:} We compare our reconstructions with model-based methods (LCT~\cite{o2018confocal}, f-k~\cite{lindell2019wave}, RSD~\cite{liu2019non}), rendering-based methods (NeTF~\cite{shen2021non}), learning-based virtual scanning (VS) method~\cite{cui2024virtual}, and optimization-based methods (CC-SOCR~\cite{liu2023non}). For fair comparisons, we interpolate any sparse measurements onto a regular grid using \texttt{scipy.interpolate.griddata} for the model-based approaches.
  For our method, the generated virtual confocal measurements are reconstructed using LCT as the backend conventional solver in Rows~1--4 and RSD in Rows~5--6.}
  \Description{The figure contains the NLOS reconstruction results for confocal configuration.}
  \label{fig:confocal-result}
\end{figure*}

\subsection{Training Process}
Based on the transient rendering pipeline in Sec.~\ref{sec:gs_transient_rendering}, we reconstruct the hidden scene by optimizing a set of 3D Gaussian primitives to match the measured transients on the relay surface (Fig.~\ref{fig:pipeline}).

We initialize $K$ Gaussian primitives parameterized by $(\boldsymbol{x}_{g,k},\boldsymbol{\Sigma}_k,\rho_k)$ (Sec.~\ref{sec:gs_transient_rendering}).
In practice, we optimize an unconstrained set of raw variables 
\[
\{\Theta_k\}_{k=1}^{K}
\triangleq
\{ \theta_{x,k},\theta_{\Sigma,k},\theta_{\rho,k}\}_{k=1}^{K}
\],
which are mapped to physically valid parameters $(\boldsymbol{x}_{g,k},\boldsymbol{\Sigma}_k,\rho_k)$. 

The center $\boldsymbol{x}_{g,k}$ is constrained to lie within a user-defined region of interest (ROI), and the albedo $\rho_k$ is non-negative. We enforce both constraints via smooth re-parameterizations (e.g., $\tanh$, softplus). 
To guarantee the positive definiteness of the covariance matrix ($\boldsymbol{\Sigma}_k\succ\boldsymbol{0}$), we parameterize it using a rotation-scaling decomposition:
$
\boldsymbol{\Sigma}_k \;=\; \boldsymbol{R}_k\,\boldsymbol{S}_k\boldsymbol{S}_k^{\top}\boldsymbol{R}_k^{\top},
$
where $\boldsymbol{S}_k=\mathrm{diag}(\boldsymbol{s}_k)$ is a diagonal scaling matrix and $\boldsymbol{R}_k\in\mathrm{SO}(3)$ is a rotation matrix comprising learnable parameters. This parameterization follows the standard practices in 3DGS, facilitating stable optimization while preserving full differentiability.

Unless otherwise specified, all primitives share the same initial isotropic covariance and albedo, while their centers are uniformly randomized within the ROI. 
The global gain $g$ (Sec.~\ref{sec:gs_transient_rendering}) is kept fixed during training and is set per dataset by scaling the rendered transient to match the overall magnitude of the measured transient at initialization.

We estimate a normal field $\{\boldsymbol{n}_p\}_{p=1}^{P}$ from the LOS point cloud $\mathcal{Q}$ using a standard point-to-normal procedure.
Specifically, for each relay point, we identify its $M$ nearest neighbors (k-NN) and compute the local normal via principal component analysis on the local neighborhood.
We then associate each measured transient with its corresponding surface normal, yielding the set $\{\boldsymbol{\tau}_{p,\boldsymbol{n}_p}\}_{p=1}^{P}$.

We optimize the parameters by minimizing the discrepancy between rendered transients $\hat{\boldsymbol{\tau}}_p$ and the measurements $\boldsymbol{\tau}_p$:
\[
\min_{\{\Theta_k\}}\;
\sum_{p=1}^{P}
\left\|\hat{\boldsymbol{\tau}}_p-\boldsymbol{\tau}_p\right\|_2^2.
\]
In practice, we use mini-batches for training to reduce sensitivity to measurement noise and outliers. We use the AdamW optimizer \cite{loshchilov2017decoupled} and assign separate learning rates to parameter groups (centers, scales, rotations, and albedo), which improves training stability by balancing updates across variables with different numerical scales. After $N$ iterations, we obtain $\{\mathcal{G}_k^{(N)}\}_{k=1}^{K}$ as the reconstructed hidden-scene representation.

To obtain a compact and visually cleaner representation, we apply a lightweight pruning step to produce a refined set $\{\mathcal{G}_k^\star\}_{k=1}^{K^\star}$ with $K^\star\le K$. Specifically, we discard primitives that (i) lie close to the ROI boundaries, or (ii) make negligible contribution due to low albedo.

\begin{figure*}[t]
  \centering
  \includegraphics[width=0.75\linewidth]{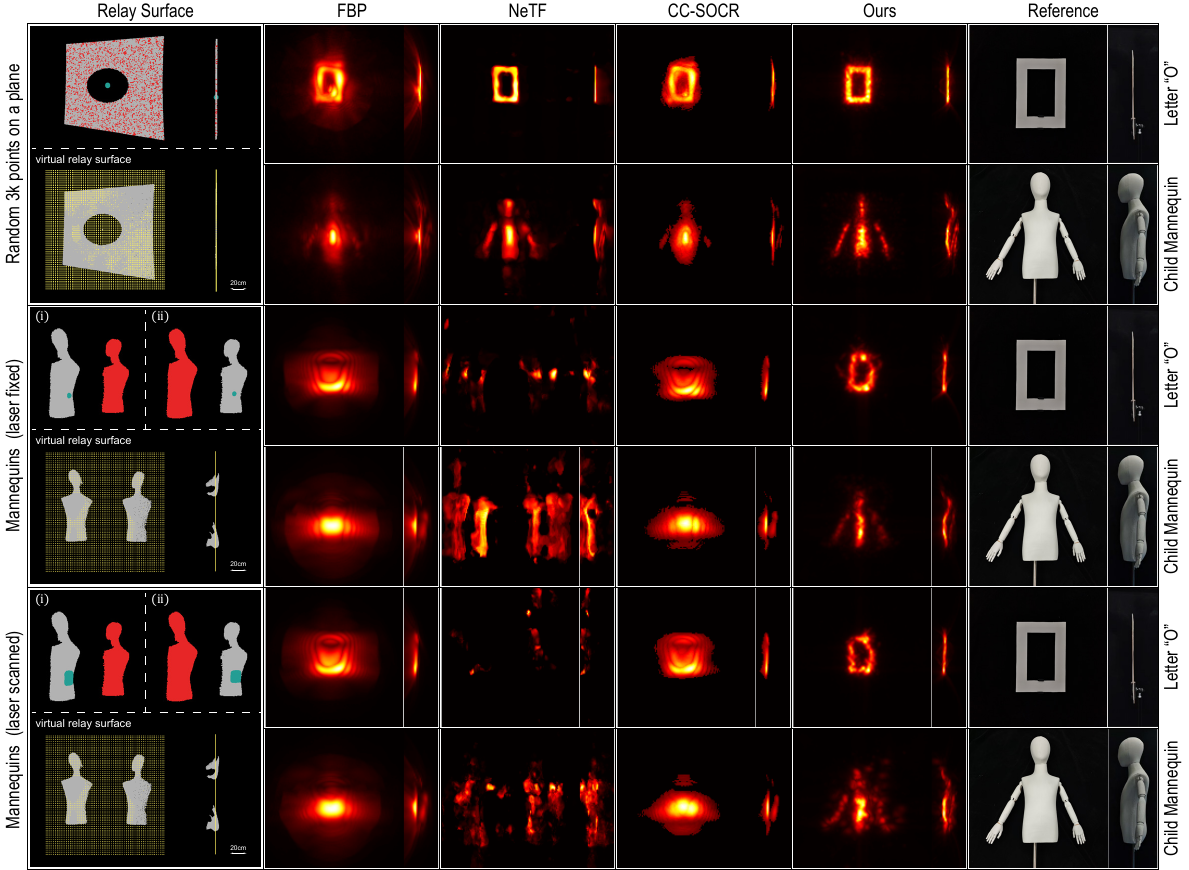}
  \caption{\textbf{Reconstruction results for non-confocal configurations.} \textbf{Data:} Hidden targets include a planar letter ``O'' ($45\,\text{cm}\times 30\,\text{cm}$, stroke width $\sim 5\,\text{cm}$) and a child mannequin (shoulder width $\sim 30\,\text{cm}$, height $\sim 70\,\text{cm}$), positioned $\sim 0.7\,\text{m}$ from the relay surface. Measurements were captured using our LOS-guided pipeline. The LOS point cloud is visualized in gray, with illumination points in green and detection points in red. The generated virtual relay surface (yellow) is displayed in both front and top views. We evaluate two relay surface configurations: a planar wall (Rows 1--2) using a fixed illumination point with 3000 random detection points; and a pair of torso mannequins, where illumination and detection points are located on separate mannequins. For the mannequin setup, we consider two regimes: a fixed laser position on each mannequin (Rows 3--4) and a locally scanned laser within a small region (Rows 5--6); see Supplementary Sec.~2 for details. \textbf{Comparison:} We compare against FBP, NeTF, and CC-SOCR, as they are capable of handling arbitrary illumination-detection pairs. All methods reconstruct a $1.5\,\text{m} \times 1.5\,\text{m} \times 1.5\,\text{m}$ hidden volume, visualized at $128 \times 128$ resolution for front views and $128 \times 32$ for side views. For our method, the generated virtual confocal measurements are synthesized   on the $z=0$ virtual relay surface and reconstructed using LCT as the backend conventional solver.
  }
  \label{fig:non-confocal-result}
  \Description{The figure contains typical non-confocal results.}
\end{figure*}
\section{Experiments}
\subsection{Custom-built Transient Measurement Setup}
To validate our approach, we build an NLOS imaging system (shown in Fig.~\ref{fig:method}) to capture high-quality transient data with different relay surface geometries. The system employs a non-coaxial two-arm configuration featuring a pair of two-axis galvanometers, which undergoes careful alignment calibration. A Time-to-Digital Converter (TDC) records the transient histograms based on the time-stamp readout signals from the SPAD. By virtue of completely independent transmit-receive architecture, our setup is well-suited for measuring transients on arbitrary relay surface geometries.
 
To characterize the relay surface, we place a diffuser in front of the laser port and simultaneously scan both the laser and SPAD ports to obtain the first-bounce transient signals. By setting proper thresholds to the depth and intensity of these signals, we automatically obtain a valid relay surface for subsequent NLOS measurements. 
With the obtained point cloud data, we can define arbitrary detection regions. See Supplementary Sec. 1 for details. 

\subsection{Confocal Results}
For confocal detection, to mitigate \textit{pile-up} effect \cite{heide2019non} arising from SPAD, we add a spatial offset between illumination and detection points. While this offset is straightforward to implement for planar walls (e.g., a uniform $10\,\mathrm{cm}$), it becomes non-trivial for complex non-planar surfaces. Thus, for confocal comparisons, we restrict our evaluation to planar relay surfaces.

To evaluate reconstruction performance under challenging sampling conditions, we benchmark the methods along three distinct aspects: \textit{region-limited}, \textit{aperture-limited}, and \textit{resolution-limited} measurements. The results are shown in Fig. \ref{fig:confocal-result}. 

\paragraph{Region-limited.} We use the \textbf{FK-Teaser} scene—a cluttered multi-object arrangement with varying depths and surface properties—to assess whether reconstruction quality can be maintained under sparse spatial sampling.  We consider two challenging sampling patterns: (i) \textbf{structured sparse sampling}, where measurements are restricted to a certain shape (the ``NLO'' letters), and (ii) \textbf{random sparse sampling} with 3,000 randomly selected points. Across both settings, our method achieves superior overall performance, simultaneously preserving the geometry of multiple objects and maintaining high reconstruction fidelity. 
\paragraph{Aperture-limited} Unlike the region-limited setting where captured signals are distributed across the relay surface, here we investigate the impact of a severely restricted measurement aperture—potentially even smaller than the hidden object itself.  We choose the \textbf{Statue} scene for this test, as it contains a single object centered within the measurement region. We progressively reduce the effective wall aperture to $1\,\mathrm{m}\times 1\,\mathrm{m}$ and $0.5\, \mathrm{m}\times 0.5\, \mathrm{m}$ (using a $16 \times 16$ grid). As illustrated, our method consistently outperforms prior methods, reconstructing a clearer statue even when the effective aperture is significantly smaller than the object. 
\paragraph{Resolution-limited.} We further evaluate the resolution limits of 3D-GTR under spatial undersampling conditions. Using our custom-built system, we capture transients of a $2\thinspace \mathrm{cm}$-resolution plate and subsample the relay surface measurements to $41 \times 41$ and $21\times 21$ grids, corresponding to $2\thinspace \mathrm{cm}$ and $4\thinspace \mathrm{cm}$ spacing respectively. Our results demonstrate that 3D-GTR exhibits a clear super-resolution effect: it successfully resolves the $2\thinspace \mathrm{cm}$ features even when the sampling pitch exceeds the target resolution. In contrast, competing methods degrade substantially under the same undersampling conditions.

\begin{figure*}[p]
  \centering
  \includegraphics[width=1.0\linewidth]{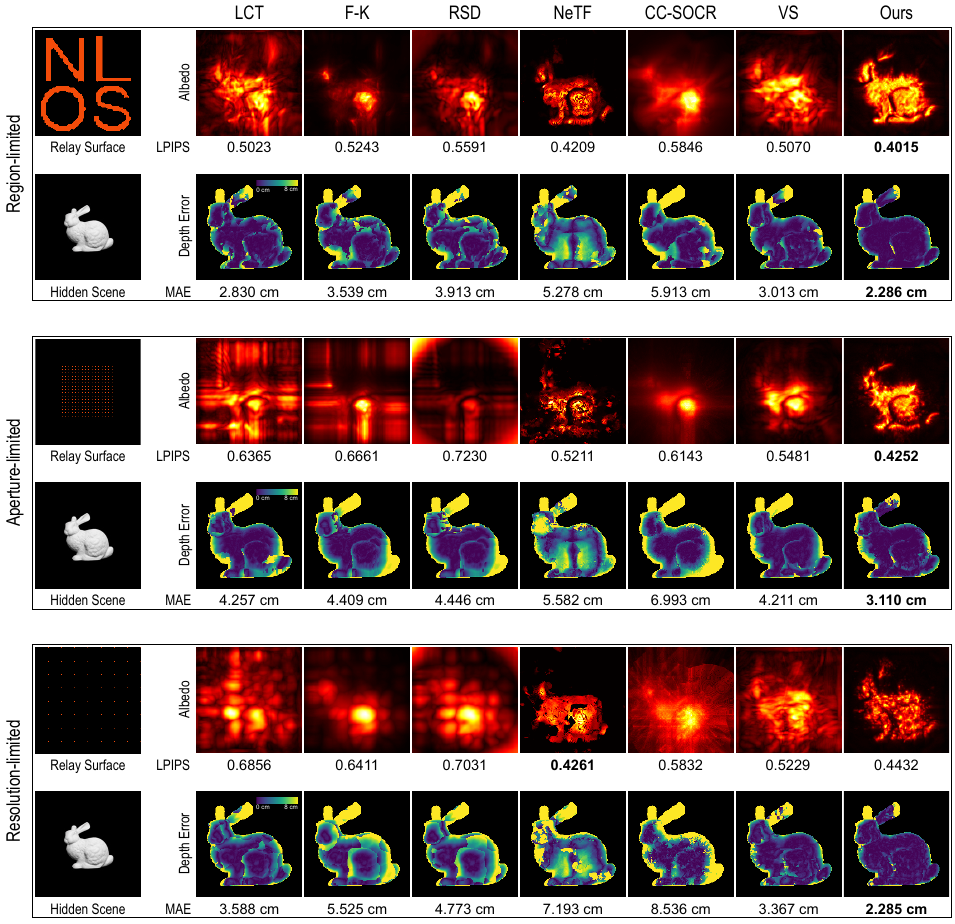}
  \caption{
\textbf{Quantitative simulation results under limited confocal measurements.}
We conduct a simulation study on the Stanford Bunny from the Zaragoza dataset~\cite{galindo19-NLOSDataset,jarabo2014framework} to quantify reconstruction accuracy under restricted relay-wall measurements.
Transient measurements are generated on a $1\,\mathrm{m}\times1\,\mathrm{m}$ planar relay wall with a dense $128\times128$ confocal scan, and the hidden-scene bounding box is centered $0.5\,\mathrm{m}$ from the wall.
We evaluate three restricted-measurement settings: \textit{region-limited} measurements using the same ``NLOS'' shaped sampling mask as in Fig.~\ref{fig:confocal-result}, \textit{aperture-limited} measurements using a centered $16\times16$ sampling region within a $0.5\,\mathrm{m}\times0.5\,\mathrm{m}$ aperture, and \textit{resolution-limited} measurements obtained by uniformly downsampling the full scan to an $8\times8$ grid.
The compared baselines and sparse-to-grid interpolation  are the same as in Fig.~\ref{fig:confocal-result}.
For each method, we report the reconstructed albedo with Learned Perceptual Image Patch Similarity (LPIPS)~\cite{zhang2018unreasonable} and the depth error map with mean absolute error (MAE).
LPIPS is computed using the AlexNet-based metric, where grayscale albedo images are replicated to three channels and normalized to $[-1,1]$; depth MAE is computed only over foreground pixels with valid ground-truth depth. For both LPIPS and MAE, lower values indicate better performance.
Our method achieves the lowest depth MAE in all three settings, while also producing competitive albedo fidelity, indicating that its robustness under limited measurements is consistent across both real and simulated data.
}
  \label{fig:simulation-result}
  \Description{The figure contains simulation results for confocal detection.}
\end{figure*}

We further quantify this behavior on a simulation study.
Figure \ref{fig:simulation-result} reports both albedo fidelity and depth accuracy under region-limited, aperture-limited, and resolution-limited confocal measurements.
The results show that our method consistently achieves the highest depth accuracy and remains competitive in albedo fidelity, supporting the robustness observed on real measurements.

\subsection{Non-confocal Results}
Compared to confocal settings, non-confocal configurations decouple the illumination and detection positions. This eliminates the geometry constraint between the illumination and detection paths, making the system more compatible with real-world deployment. 
The comparison results are illustrated in Fig. \ref{fig:non-confocal-result}. In the planar setting, both NeTF and our approach outperform the other methods. However, the finer details of the letter ``O'' illustrate that NeTF produces slight distortions near the lower-left corner, while our method correctly recovers the rectangular shape matching the ground truth. Regarding the arbitrary relay surface, our approach substantially outperforms all baselines. Under the laser-fixed acquisition configuration, the reconstructed mannequin head is partially missing, likely due to insufficient illumination coverage. To capture additional information, we further scan the illumination over a small region to introduce viewpoint diversity. As shown in Fig. \ref{fig:non-confocal-result}, this enables recovery of the missing head region, although it introduces mild arm distortions. We attribute the distortions to increased sensitivity to measurement errors and temporal fluctuations when the illumination position is variable. 

\begin{figure*}[p]
  \centering
  \includegraphics[width=0.9\linewidth]{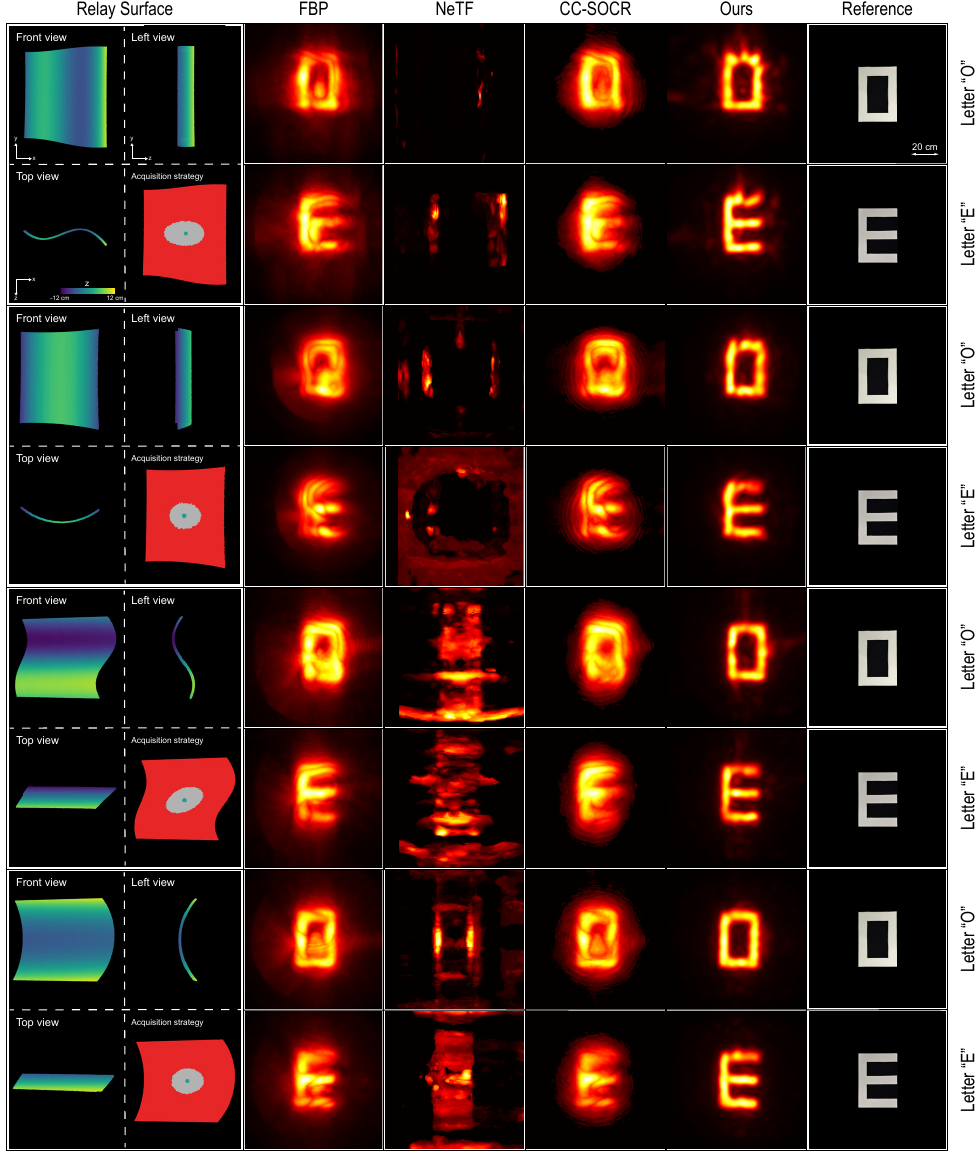}
 \caption{
\textbf{Non-confocal reconstruction with diverse relay surface geometries.}
We evaluate NLOS reconstruction on real measurements captured by our custom-built system under four non-planar relay surface geometries: horizontally undulating S-shaped and C-shaped surfaces, and vertically undulating S-shaped and C-shaped surfaces.
For each relay surface, we reconstruct two hidden planar targets, the letters ``O'' and ``E'', placed approximately $0.7\,\mathrm{m}$ from the relay surface.
The left column visualizes each relay surface from front, left, and top views, together with the acquisition strategy used for NLOS detection, where the green point denotes the fixed laser position and the red region denotes the scanned detection positions.
We compare against the same non-confocal baselines as in Fig.~\ref{fig:non-confocal-result}, including FBP, NeTF, and CC-SOCR.
All methods reconstruct a hidden volume with a lateral field of view of $1\,\mathrm{m}\times1\,\mathrm{m}$.
For our method, the virtual relay surface is set to be the $z=0$ plane, and RSD is used as the backend solver.
Across these curved relay surfaces, our method preserves the target letter shapes more reliably than the competing methods, supporting its robustness to arbitrary relay surface geometries.}
  \label{fig:nonplanar-result}
  \Description{The figure contains diverse non-planar results.}
\end{figure*}

We also evaluate our method on a broader set of non-planar relay surfaces, including curved surfaces with horizontal and vertical undulations, as shown in Fig.~\ref{fig:nonplanar-result}.
Our method consistently recovers clearer target shapes than the non-confocal baselines across these geometries.
This supports the generalization ability of 3D-GTR under arbitrary relay surface geometries.

\subsection{Runtime Analysis}
We benchmark reconstruction speed on a workstation equipped with an Intel Xeon Platinum 8352V (36 cores, 2.10 GHz) CPU and an NVIDIA RTX A6000 (48 GB) GPU. We focus on methods that can handle arbitrary transients, specifically FBP, NeTF, CC-SOCR, and our approach. All reported runtimes are representative values from the experiments in Fig. \ref{fig:non-confocal-result}. As shown in Table~\ref{tab:runtime_analysis}, our method demonstrates a significant efficiency advantage. During the optimization/training stage, our differentiable rendering pipeline converges substantially faster than NeTF. During inference, the highly parallel nature of 3D Gaussian primitives enables extremely rapid evaluation compared to other methods. Notably, the reported inference time for our method also includes the runtime of a conventional transient solver (LCT). Additional analysis is shown in Supplementary Sec. 3.
\begin{table}[htpb]
  \centering
  \caption{Reconstruction time of different methods. The values are the average reconstruction time for a $1.5\,\mathrm{m}\times1.5\,\mathrm{m}\times1.5\,\mathrm{m}$ volume using a training dataset of $\sim 8000$ transients.}
  \label{tab:runtime_analysis}

  \setlength{\tabcolsep}{4.5pt}
  \begin{tabular}{@{}lcccc@{}}
    \toprule
    & FBP & CC-SOCR & NeTF & Ours \\
    \midrule
    \shortstack[l]{Training time}  & N/A    & N/A        & 0.31 h/epoch & 10.9 s/epoch \\
    \shortstack[l]{Inference time} & 25.9 s & 1.1 h/loop & 4.2 s        & 0.6 s \\
    \bottomrule
  \end{tabular}
\end{table}

\subsection{Ablation Study}
\begin{figure}[h]
  \centering  \includegraphics[width=0.9\linewidth]{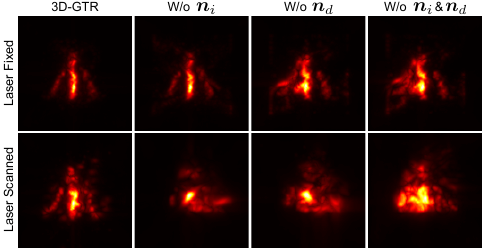}
  \caption{\textbf{Ablation study on geometry attributes in rendering model.}
  Reconstruction of child mannequin, using full 3D-GTR model, model without $\boldsymbol{n}_i$, without $\boldsymbol{n}_d$, and without $\boldsymbol{n}_i$ and $\boldsymbol{n}_d$.} 
  \label{fig:ablation}
  \Description{The figure contains the ablation study on geometry attributes in rendering model}
\end{figure}
One key factor behind our superior performance is the \textbf{geometry-aware transient rendering} formulation in Eq. \eqref{eq:geometry}. To assess the individual impact of these two terms, we conduct an ablation study in the non-confocal setting on mannequin-based (arbitrary) relay surfaces with illumination points fixed and changed, using the child mannequin as the hidden target (see Fig. \ref{fig:ablation}). In the laser-fixed scenario, removing illumination normals $\boldsymbol{n}_i$ has only a minor effect, as $\boldsymbol{n}_i$ remains constant during detection on a single mannequin. In contrast, in all remaining settings, omitting $\boldsymbol{n}_i$ and/or $\boldsymbol{n}_d$ induces a pronounced model mismatch, as these normals vary substantially across the sampled points, leading to clearly degraded results. Explicitly accounting for local surface orientation yields a more faithful approximation to real light-transport physics, which in turn leads to consistently improved reconstruction quality. 

\section{Limitations and Conclusions}
Similar to NeTF and CC-SOCR, 3D-GTR is scene-specific and therefore requires reoptimization when the hidden scene changes. While our method is substantially faster than these baselines, it does not yet achieve real-time performance. Nevertheless, inspired by the original 3DGS formulation, we find that good initialization of Gaussian primitives can significantly accelerate convergence. Moreover, because our representation is physically grounded and differentiable, it can be naturally integrated with advanced video reconstruction techniques, such as plug-and-play priors \cite{ye2024plug} or spatiotemporal transformers \cite{li2025transit}, offering promising pathways toward real-time NLOS imaging. 

In summary, we establish a novel NLOS imaging framework without assumptions on relay surface geometry. By representing the hidden scene using 3D Gaussian primitives and leveraging the geometry information captured by LOS measurements, we introduce 3D-GTR as a fully differentiable transient rendering model that reformulates hidden scene reconstruction as an end-to-end optimization from raw NLOS transients. To enable reconstruction in a user-specified space, we feed the optimized Gaussian representations back into the renderer to synthesize dense, regularly sampled confocal transients, which can then be processed by standard NLOS solvers.

Extensive real-world experiments demonstrate that our approach is substantially more efficient than existing methods while delivering higher fidelity and more reliable reconstructions across confocal and non-confocal settings, and for both planar and non-planar relay surfaces. We attribute this performance superiority to three key factors.
First, because 3D Gaussian primitives are lightweight and sparse, they avoid redundant computation inherent to voxel- or mesh-based representations. Moreover, the highly parallel structure of Gaussian rendering makes our pipeline better suited to mainstream GPU architectures than implicit-neural representations. Second, the physically grounded parameterization allows the optimization process to partially mitigate practical imperfections—such as relay surface geometry errors, measurement noise, and detector time jitter—by adjusting the position, orientation, scale, and albedo of the primitives. Third, our explicit formulation of light transport, accounting for surface normals, extracts more informative cues from ambiguous real-world measurements, leading to marked enhancement on complex, irregular, and arbitrary relay surface geometries.  

Finally, NLOS imaging holds transformative potential for applications ranging from autonomous navigation to public safety. With rapid advances in single-photon detection and high-performance computing, practical NLOS imaging is increasingly within reach. Our work takes a concrete step toward promoting NLOS imaging from controlled laboratory settings to real-world deployment by shifting geometry-assumed pipelines toward a measurement-driven framework.
\begin{acks}

We acknowledge the
support of Fundamental and Interdisciplinary Disciplines Breakthrough Plan of the Ministry of Education of China (JYB2025XDXM121). 
\end{acks}

\bibliographystyle{ACM-Reference-Format}
\bibliography{reference}

%% file: reference.bib
@String{Computing = "Computing" }

@String{Computer = "{IEEE} Computer" }

@String{Springer = "Springer-Verlag" }

@inproceedings{loshchilov2017decoupled,
  author       = {Ilya Loshchilov and
                  Frank Hutter},
  title        = {Decoupled Weight Decay Regularization},
  booktitle    = {7th International Conference on Learning Representations (ICLR),
                  New Orleans, LA, USA, May 6-9, 2019},
  publisher    = {OpenReview.net},
  year         = {2019},
  url          = {https://openreview.net/forum?id=Bkg6RiCqY7},
  bibsource    = {dblp computer science bibliography, https://dblp.org}
}

@InProceedings{li2025transit,
    author    = {Li, Ruiqian and Shen, Siyuan and Xia, Suan and Wang, Ziheng and Peng, Xingyue and Song, Chengxuan and Zhu, Yingsheng and Wu, Tao and Li, Shiying and Yu, Jingyi},
    title     = {TransiT: Transient Transformer for Non-line-of-sight Videography},
    booktitle = {Proceedings of the IEEE/CVF International Conference on Computer Vision (ICCV)},
    month     = {October},
    year      = {2025},
    pages     = {27542-27551}
}

@InProceedings{scheiner2020seeing,
author = {Scheiner, Nicolas and Kraus, Florian and Wei, Fangyin and Phan, Buu and Mannan, Fahim and Appenrodt, Nils and Ritter, Werner and Dickmann, Jurgen and Dietmayer, Klaus and Sick, Bernhard and Heide, Felix},
title = {Seeing Around Street Corners: Non-Line-of-Sight Detection and Tracking In-the-Wild Using Doppler Radar},
booktitle = {Proceedings of the IEEE/CVF Conference on Computer Vision and Pattern Recognition (CVPR)},
month = {June},
pages = {2065–2074},
year = {2020}
}

@ARTICLE{rapp2020advances,
  author={Rapp, Joshua and Tachella, Julian and Altmann, Yoann and McLaughlin, Stephen and Goyal, Vivek K},
  journal={IEEE Signal Processing Magazine}, 
  title={Advances in Single-Photon Lidar for Autonomous Vehicles: Working Principles, Challenges, and Recent Advances}, 
  year={2020},
  volume={37},
  number={4},
  pages={62-71},
  doi={10.1109/MSP.2020.2983772}}

@article{buttafava2015non,
author = {Mauro Buttafava and Jessica Zeman and Alberto Tosi and Kevin Eliceiri and Andreas Velten},
journal = {Opt. Express},
number = {16},
pages = {20997--21011},
publisher = {Optica Publishing Group},
title = {Non-line-of-sight imaging using a time-gated single photon avalanche diode},
volume = {23},
month = {Aug},
year = {2015},
url = {https://opg.optica.org/oe/abstract.cfm?URI=oe-23-16-20997},
doi = {10.1364/OE.23.020997},
}

@INPROCEEDINGS{kirmani2009looking,
  author={Kirmani, Ahmed and Hutchison, Tyler and Davis, James and Raskar, Ramesh},
  booktitle={2009 IEEE 12th International Conference on Computer Vision}, 
  title={Looking around the corner using transient imaging}, 
  year={2009},
  volume={},
  number={},
  pages={159-166},
  doi={10.1109/ICCV.2009.5459160}}

@article{velten2012recovering,
   author = {Velten, Andreas and Willwacher, Thomas and Gupta, Otkrist and Veeraraghavan, Ashok and Bawendi, Moungi G. and Raskar, Ramesh},
   title = {Recovering three-dimensional shape around a corner using ultrafast time-of-flight imaging},
   journal = {Nature Communications},
   volume = {3},
   number = {1},
   pages = {745},
   ISSN = {2041-1723},
   DOI = {10.1038/ncomms1747},
   url = {https://doi.org/10.1038/ncomms1747},
   year = {2012},
   type = {Journal Article}
}

@article{o2018confocal,
   author = {O’Toole, Matthew and Lindell, David B. and Wetzstein, Gordon},
   title = {Confocal non-line-of-sight imaging based on the light-cone transform},
   journal = {Nature},
   volume = {555},
   number = {7696},
   pages = {338-341},
   ISSN = {1476-4687},
   DOI = {10.1038/nature25489},
   url = {https://doi.org/10.1038/nature25489},
   year = {2018},
   type = {Journal Article}
}

@article{lindell2019wave,
author = {Lindell, David B. and Wetzstein, Gordon and O'Toole, Matthew},
title = {Wave-based non-line-of-sight imaging using fast f-k migration},
year = {2019},
issue_date = {August 2019},
publisher = {Association for Computing Machinery},
address = {New York, NY, USA},
volume = {38},
number = {4},
issn = {0730-0301},
url = {https://doi.org/10.1145/3306346.3322937},
doi = {10.1145/3306346.3322937},
journal = {ACM Trans. Graph.},
month = jul,
articleno = {116},
numpages = {13}
}

@article{liu2019non,
   author = {Liu, Xiaochun and Guillén, Ibón and La Manna, Marco and Nam, Ji Hyun and Reza, Syed Azer and Huu Le, Toan and Jarabo, Adrian and Gutierrez, Diego and Velten, Andreas},
   title = {Non-line-of-sight imaging using phasor-field virtual wave optics},
   journal = {Nature},
   volume = {572},
   number = {7771},
   pages = {620-623},
   ISSN = {1476-4687},
   DOI = {10.1038/s41586-019-1461-3},
   url = {https://doi.org/10.1038/s41586-019-1461-3},
   year = {2019},
   type = {Journal Article}
}

@article{liu2020phasor,
   author = {Liu, Xiaochun and Bauer, Sebastian and Velten, Andreas},
   title = {Phasor field diffraction based reconstruction for fast non-line-of-sight imaging systems},
   journal = {Nature Communications},
   volume = {11},
   number = {1},
   pages = {1645},
   ISSN = {2041-1723},
   DOI = {10.1038/s41467-020-15157-4},
   url = {https://doi.org/10.1038/s41467-020-15157-4},
   year = {2020},
   type = {Journal Article}
}

@article{zhang2025sub,
author = {Wenjun Zhang and Shuo Zhu and Lijia Chen and Lingfeng Liu and Lianfa Bai and Edmund Y. Lam and Enlai Guo and Jing Han},
journal = {Opt. Express},
number = {14},
pages = {30783--30798},
publisher = {Optica Publishing Group},
title = {Sub-pixel resolving modulation for non-line-of-sight imaging},
volume = {33},
month = {Jul},
year = {2025},
url = {https://opg.optica.org/oe/abstract.cfm?URI=oe-33-14-30783},
doi = {10.1364/OE.569102},
}

@ARTICLE{la2018error,
  author={La Manna, Marco and Kine, Fiona and Breitbach, Eric and Jackson, Jonathan and Sultan, Talha and Velten, Andreas},
  journal={IEEE Transactions on Pattern Analysis and Machine Intelligence}, 
  title={Error Backprojection Algorithms for Non-Line-of-Sight Imaging}, 
  year={2019},
  volume={41},
  number={7},
  pages={1615-1626},
  doi={10.1109/TPAMI.2018.2843363}}

@article{li2022fast,
author = {Zhupeng Li and Xintong Liu and Jianyu Wang and Zuoqiang Shi and Lingyun Qiu and Xing Fu},
journal = {Opt. Lett.},
number = {8},
pages = {1928--1931},
publisher = {Optica Publishing Group},
title = {Fast non-line-of-sight imaging based on first photon event stamping},
volume = {47},
month = {Apr},
year = {2022},
url = {https://opg.optica.org/ol/abstract.cfm?URI=ol-47-8-1928},
doi = {10.1364/OL.446079},
}

@InProceedings{yu2023enhancing,
    author    = {Yu, Yanhua and Shen, Siyuan and Wang, Zi and Huang, Binbin and Wang, Yuehan and Peng, Xingyue and Xia, Suan and Liu, Ping and Li, Ruiqian and Li, Shiying},
    title     = {Enhancing Non-line-of-sight Imaging via Learnable Inverse Kernel and Attention Mechanisms},
    booktitle = {Proceedings of the IEEE/CVF International Conference on Computer Vision (ICCV)},
    month     = {October},
    year      = {2023},
    pages     = {10563-10573}
}

@misc{wei2025fast,
Author = {Yijun Wei and Jianyu Wang and Leping Xiao and Zuoqiang Shi and Xing Fu and Lingyun Qiu},
Title = {Fast and Memory-efficient Non-line-of-sight Imaging with Quasi-Fresnel Transform},
Year = {2025},
Eprint = {2508.02003},
}

@inproceedings{li2023deep,
 author = {Li, Yue and Zhang, Yueyi and Ye, Juntian and Xu, Feihu and Xiong, Zhiwei},
 booktitle = {Advances in Neural Information Processing Systems},
 editor = {A. Oh and T. Naumann and A. Globerson and K. Saenko and M. Hardt and S. Levine},
 pages = {59095--59106},
 publisher = {Curran Associates, Inc.},
 title = {Deep Non-line-of-sight Imaging from Under-scanning Measurements},
 url = {https://proceedings.neurips.cc/paper_files/paper/2023/file/b91cc0a242e6518ee731f74e82b2eebd-Paper-Conference.pdf},
 volume = {36},
 year = {2023}
}

@inproceedings{cui2024virtual,
 author = {Cui, Xingyu and Yue, Huanjing and Li, Song and Yin, Xiangjun and Hou, Yusen and Meng, Yun and Zou, Kai and Hu, Xiaolong and Yang, Jingyu},
 booktitle = {Advances in Neural Information Processing Systems},
 doi = {10.52202/079017-3472},
 editor = {A. Globerson and L. Mackey and D. Belgrave and A. Fan and U. Paquet and J. Tomczak and C. Zhang},
 pages = {109381--109406},
 publisher = {Curran Associates, Inc.},
 title = {Virtual Scanning: Unsupervised Non-line-of-sight Imaging from Irregularly Undersampled Transients},
 url = {https://proceedings.neurips.cc/paper_files/paper/2024/file/c58437945392cec01e0c75ff6cef901a-Paper-Conference.pdf},
 volume = {37},
 year = {2024}
}

@INPROCEEDINGS{gu2023fast,
  author={Gu, Chaoying and Sultan, Talha and Masumnia-Bisheh, Khadijeh and Waller, Laura and Velten, Andreas},
  booktitle={2023 IEEE International Conference on Computational Photography (ICCP)}, 
  title={Fast Non-line-of-sight Imaging with Non-planar Relay Surfaces}, 
  year={2023},
  volume={},
  number={},
  pages={1-12},
  doi={10.1109/ICCP56744.2023.10233262}}

@article{liu2023non,
   author = {Liu, Xintong and Wang, Jianyu and Xiao, Leping and Shi, Zuoqiang and Fu, Xing and Qiu, Lingyun},
   title = {Non-line-of-sight imaging with arbitrary illumination and detection pattern},
   journal = {Nature Communications},
   volume = {14},
   number = {1},
   pages = {3230},
   ISSN = {2041-1723},
   DOI = {10.1038/s41467-023-38898-4},
   url = {https://doi.org/10.1038/s41467-023-38898-4},
   year = {2023},
   type = {Journal Article}
}

@ARTICLE{cui2025transdiff,
  author={Cui, Xingyu and Yue, Huanjing and Sun, Shida and Li, Yue and Hou, Yusen and Xiong, Zhiwei and Yang, Jingyu},
  journal={IEEE Transactions on Image Processing}, 
  title={TransDiff: Unsupervised Non-Line-of-Sight Imaging With Aperture-Limited Relay Surfaces}, 
  year={2025},
  volume={34},
  number={},
  pages={8018-8031},
  doi={10.1109/TIP.2025.3637694}}

@article{la2020non,
author = {Marco La Manna and Ji-Hyun Nam and Syed Azer Reza and Andreas Velten},
journal = {Opt. Express},
number = {4},
pages = {5331--5339},
publisher = {Optica Publishing Group},
title = {Non-line-of-sight-imaging using dynamic relay surfaces},
volume = {28},
month = {Feb},
year = {2020},
url = {https://opg.optica.org/oe/abstract.cfm?URI=oe-28-4-5331},
doi = {10.1364/OE.383586},
}

@article{ye2021compressed,
author = {Jun-Tian Ye and Xin Huang and Zheng-Ping Li and Feihu Xu},
journal = {Opt. Express},
number = {2},
pages = {1749--1763},
publisher = {Optica Publishing Group},
title = {Compressed sensing for active non-line-of-sight imaging},
volume = {29},
month = {Jan},
year = {2021},
url = {https://opg.optica.org/oe/abstract.cfm?URI=oe-29-2-1749},
doi = {10.1364/OE.413774},
}

@InProceedings{wang2023non,
    author    = {Wang, Jianyu and Liu, Xintong and Xiao, Leping and Shi, Zuoqiang and Qiu, Lingyun and Fu, Xing},
    title     = {Non-Line-of-Sight Imaging With Signal Superresolution Network},
    booktitle = {Proceedings of the IEEE/CVF Conference on Computer Vision and Pattern Recognition (CVPR)},
    month     = {June},
    year      = {2023},
    pages     = {17420-17429}
}

@ARTICLE{metzler2020keyhole,
  author={Metzler, Christopher A. and Lindell, David B. and Wetzstein, Gordon},
  journal={IEEE Transactions on Computational Imaging}, 
  title={Keyhole Imaging: Non-Line-of-Sight Imaging and Tracking of Moving Objects Along a Single Optical Path}, 
  year={2021},
  volume={7},
  number={},
  pages={1-12},
  doi={10.1109/TCI.2020.3046472}}

@article{heide2019non,
author = {Heide, Felix and O’Toole, Matthew and Zang, Kai and Lindell, David B. and Diamond, Steven and Wetzstein, Gordon},
title = {Non-line-of-sight Imaging with Partial Occluders and Surface Normals},
year = {2019},
issue_date = {June 2019},
publisher = {Association for Computing Machinery},
address = {New York, NY, USA},
volume = {38},
number = {3},
issn = {0730-0301},
url = {https://doi.org/10.1145/3269977},
doi = {10.1145/3269977},
journal = {ACM Trans. Graph.},
month = may,
articleno = {22},
numpages = {10}
}

@article{miao2025under,
    author = {Miao, Jinye and Shi, Yingjie and Liu, Lingfeng and Wei, Yi and Cai, Fuyao and Bai, Lianfa and Guo, Enlai and Han, Jing},
    title = {Under-scanning non-line-of-sight imaging based on convolution approximation and optimization},
    journal = {APL Photonics},
    volume = {10},
    number = {6},
    pages = {066110},
    year = {2025},
    month = {June},
    issn = {2378-0967},
    doi = {10.1063/5.0266391},
    url = {https://doi.org/10.1063/5.0266391},
}

@InProceedings{liu2023few,
    author    = {Liu, Xintong and Wang, Jianyu and Xiao, Leping and Fu, Xing and Qiu, Lingyun and Shi, Zuoqiang},
    title     = {Few-Shot Non-Line-of-Sight Imaging With Signal-Surface Collaborative Regularization},
    booktitle = {Proceedings of the IEEE/CVF Conference on Computer Vision and Pattern Recognition (CVPR)},
    month     = {June},
    year      = {2023},
    pages     = {13303-13312}
}

@article{ye2024plug,
author = {Ye, Juntian and Hong, Yu and Su, Xiongfei and Yuan, Xin and Xu, Feihu},
title = {Plug-and-Play Algorithms for Dynamic Non-line-of-sight Imaging},
year = {2024},
issue_date = {October 2024},
publisher = {Association for Computing Machinery},
address = {New York, NY, USA},
volume = {43},
number = {5},
issn = {0730-0301},
url = {https://doi.org/10.1145/3665139},
doi = {10.1145/3665139},
journal = {ACM Trans. Graph.},
month = jun,
articleno = {155},
numpages = {12}
}

@article{ye2024real,
   author = {Ye, Jun-Tian and Sun, Yi and Li, Wenwen and Zeng, Jian-Wei and Hong, Yu and Li, Zheng-Ping and Huang, Xin and Xue, Xianghui and Yuan, Xin and Xu, Feihu and Dou, Xiankang and Pan, Jian-Wei},
   title = {Real-time non-line-of-sight computational imaging using spectrum filtering and motion compensation},
   journal = {Nature Computational Science},
   volume = {4},
   number = {12},
   pages = {920-927},
   ISSN = {2662-8457},
   DOI = {10.1038/s43588-024-00722-4},
   url = {https://doi.org/10.1038/s43588-024-00722-4},
   year = {2024},
   type = {Journal Article}
}

@inproceedings{li2024toward,
 author = {Li, Yue and Sun, Yi and Sun, Shida and Ye, Juntian and Zhang, Yueyi and Xu, Feihu and Xiong, Zhiwei},
 booktitle = {Advances in Neural Information Processing Systems},
 doi = {10.52202/079017-4016},
 editor = {A. Globerson and L. Mackey and D. Belgrave and A. Fan and U. Paquet and J. Tomczak and C. Zhang},
 pages = {126452--126473},
 publisher = {Curran Associates, Inc.},
 title = {Toward Dynamic Non-Line-of-Sight Imaging with Mamba Enforced Temporal Consistency},
 url = {https://proceedings.neurips.cc/paper_files/paper/2024/file/e481829e70a46db98c0c2eb46ff91bac-Paper-Conference.pdf},
 volume = {37},
 year = {2024}
}

@article{sun2026cuda,
title = {CUDA-accelerated Non-line-of-sight imaging with irregular relay surfaces},
journal = {Optics and Lasers in Engineering},
volume = {200},
pages = {109591},
year = {2026},
issn = {0143-8166},
doi = {https://doi.org/10.1016/j.optlaseng.2025.109591},
url = {https://www.sciencedirect.com/science/article/pii/S0143816625007754},
author = {Yi Sun and Yu Hong and Ziheng Qiu and Wei Li and Wenwen Li and Qilin Sun and Feihu Xu}
}

@article{yi2021differentiable,
author = {Yi, Shinyoung and Kim, Donggun and Choi, Kiseok and Jarabo, Adrian and Gutierrez, Diego and Kim, Min H.},
title = {Differentiable transient rendering},
year = {2021},
issue_date = {December 2021},
publisher = {Association for Computing Machinery},
address = {New York, NY, USA},
volume = {40},
number = {6},
issn = {0730-0301},
url = {https://doi.org/10.1145/3478513.3480498},
doi = {10.1145/3478513.3480498},
journal = {ACM Trans. Graph.},
month = dec,
articleno = {286},
numpages = {11}
}

@article{wu2021differentiable,
author = {Wu, Lifan and Cai, Guangyan and Ramamoorthi, Ravi and Zhao, Shuang},
title = {Differentiable time-gated rendering},
year = {2021},
issue_date = {December 2021},
publisher = {Association for Computing Machinery},
address = {New York, NY, USA},
volume = {40},
number = {6},
issn = {0730-0301},
url = {https://doi.org/10.1145/3478513.3480489},
doi = {10.1145/3478513.3480489},
journal = {ACM Trans. Graph.},
month = dec,
articleno = {287},
numpages = {16}
}

@InProceedings{shim2024domain,
author="Shim, Hyunbo
and Cho, In
and Kwon, Daekyu
and Kim, Seon Joo",
editor="Leonardis, Ale{\v{s}}
and Ricci, Elisa
and Roth, Stefan
and Russakovsky, Olga
and Sattler, Torsten
and Varol, G{\"u}l",
title="Domain Reduction Strategy for Non-Line-of-Sight Imaging",
booktitle="Computer Vision -- ECCV 2024",
year="2025",
publisher="Springer Nature Switzerland",
address="Cham",
pages="75--92",
isbn="978-3-031-72751-1"
}

@ARTICLE{shen2024holi,
  author={Shen, Siyuan and Xia, Suan and Peng, Xingyue and Wang, Ziyu and Zhu, Yingsheng and Li, Shiying and Yu, Jingyi},
  journal={IEEE Transactions on Pattern Analysis and Machine Intelligence}, 
  title={HOLI-1-to-3: Transient-Enhanced Holistic Image-to-3D Generation}, 
  year={2025},
  volume={47},
  number={9},
  pages={7206-7217},
  doi={10.1109/TPAMI.2024.3463875}}

@inproceedings{choi2023self,
author = {Choi, Kiseok and Kim, Inchul and Choi, Dongyoung and Marco, Julio and Gutierrez, Diego and Kim, Min H.},
title = {Self-Calibrating, Fully Differentiable NLOS Inverse Rendering},
year = {2023},
isbn = {9798400703157},
publisher = {Association for Computing Machinery},
address = {New York, NY, USA},
url = {https://doi.org/10.1145/3610548.3618140},
doi = {10.1145/3610548.3618140},
booktitle = {SIGGRAPH Asia 2023 Conference Papers},
articleno = {108},
numpages = {11},
location = {Sydney, NSW, Australia},
series = {SA '23}
}

@InProceedings{plack2023fast,
    author    = {Plack, Markus and Callenberg, Clara and Schneider, Monika and Hullin, Matthias B.},
    title     = {Fast Differentiable Transient Rendering for Non-Line-of-Sight Reconstruction},
    booktitle = {Proceedings of the IEEE/CVF Winter Conference on Applications of Computer Vision (WACV)},
    month     = {January},
    year      = {2023},
    pages     = {3067-3076}
}

@InProceedings{fujimura2023nlos,
    author    = {Fujimura, Yuki and Kushida, Takahiro and Funatomi, Takuya and Mukaigawa, Yasuhiro},
    title     = {NLOS-NeuS: Non-line-of-sight Neural Implicit Surface},
    booktitle = {Proceedings of the IEEE/CVF International Conference on Computer Vision (ICCV)},
    month     = {October},
    year      = {2023},
    pages     = {10532-10541}
}

@InProceedings{tsai2019beyond,
author = {Tsai, Chia-Yin and Sankaranarayanan, Aswin C. and Gkioulekas, Ioannis},
title = {Beyond Volumetric Albedo -- A Surface Optimization Framework for Non-Line-Of-Sight Imaging},
booktitle = {Proceedings of the IEEE/CVF Conference on Computer Vision and Pattern Recognition (CVPR)},
month = {June},
pages={1545-1555},
year = {2019}
}

@article{iseringhausen2020non,
author = {Iseringhausen, Julian and Hullin, Matthias B.},
title = {Non-line-of-sight Reconstruction Using Efficient Transient Rendering},
year = {2020},
issue_date = {February 2020},
publisher = {Association for Computing Machinery},
address = {New York, NY, USA},
volume = {39},
number = {1},
issn = {0730-0301},
url = {https://doi.org/10.1145/3368314},
doi = {10.1145/3368314},
journal = {ACM Trans. Graph.},
month = jan,
articleno = {8},
numpages = {14}
}

@ARTICLE{shen2021non,
  author={Shen, Siyuan and Wang, Zi and Liu, Ping and Pan, Zhengqing and Li, Ruiqian and Gao, Tian and Li, Shiying and Yu, Jingyi},
  journal={IEEE Transactions on Pattern Analysis and Machine Intelligence}, 
  title={Non-line-of-Sight Imaging via Neural Transient Fields}, 
  year={2021},
  volume={43},
  number={7},
  pages={2257-2268},
  doi={10.1109/TPAMI.2021.3076062}}

@article{jarabo2014framework,
author = {Jarabo, Adrian and Marco, Julio and Mu\~{n}oz, Adolfo and Buisan, Raul and Jarosz, Wojciech and Gutierrez, Diego},
title = {A framework for transient rendering},
year = {2014},
issue_date = {November 2014},
publisher = {Association for Computing Machinery},
address = {New York, NY, USA},
volume = {33},
number = {6},
issn = {0730-0301},
url = {https://doi.org/10.1145/2661229.2661251},
doi = {10.1145/2661229.2661251},
journal = {ACM Trans. Graph.},
month = nov,
articleno = {177},
numpages = {10}
}

@misc{huang2023omni,
Author = {Binbin Huang and Xingyue Peng and Siyuan Shen and Suan Xia and Ruiqian Li and Yanhua Yu and Yuehan Wang and Shenghua Gao and Wenzheng Chen and Shiying Li and Jingyi Yu},
Title = {Omni-Line-of-Sight Imaging for Holistic Shape Reconstruction},
Year = {2023},
Eprint = {2304.10780},
}

@misc{grau2022occlusion,
Author = {Javier Grau and Markus Plack and Patrick Haehn and Michael Weinmann and Matthias Hullin},
Title = {Occlusion Fields: An Implicit Representation for Non-Line-of-Sight Surface Reconstruction},
Year = {2022},
Eprint = {2203.08657},
}

@ARTICLE{mu2022physics,
  author={Mu, Fangzhou and Mo, Sicheng and Peng, Jiayong and Liu, Xiaochun and Nam, Ji Hyun and Raghavan, Siddeshwar and Velten, Andreas and Li, Yin},
  journal={IEEE Transactions on Pattern Analysis and Machine Intelligence}, 
  title={Physics to the Rescue: Deep Non-Line-of-Sight Reconstruction for High-Speed Imaging}, 
  year={2025},
  volume={47},
  number={8},
  pages={6146-6158},
  doi={10.1109/TPAMI.2022.3203383}}

@article{pediredla2019ellipsoidal,
author = {Pediredla, Adithya and Veeraraghavan, Ashok and Gkioulekas, Ioannis},
title = {Ellipsoidal path connections for time-gated rendering},
year = {2019},
issue_date = {August 2019},
publisher = {Association for Computing Machinery},
address = {New York, NY, USA},
volume = {38},
number = {4},
issn = {0730-0301},
url = {https://doi.org/10.1145/3306346.3323016},
doi = {10.1145/3306346.3323016},
journal = {ACM Trans. Graph.},
month = jul,
articleno = {38},
numpages = {12}
}

@article{chen2020learned,
author = {Chen, Wenzheng and Wei, Fangyin and Kutulakos, Kiriakos N. and Rusinkiewicz, Szymon and Heide, Felix},
title = {Learned feature embeddings for non-line-of-sight imaging and recognition},
year = {2020},
issue_date = {December 2020},
publisher = {Association for Computing Machinery},
address = {New York, NY, USA},
volume = {39},
number = {6},
issn = {0730-0301},
url = {https://doi.org/10.1145/3414685.3417825},
doi = {10.1145/3414685.3417825},
journal = {ACM Trans. Graph.},
month = nov,
articleno = {230},
numpages = {18}
}

@article{kerbl20233d,
author = {Kerbl, Bernhard and Kopanas, Georgios and Leimkuehler, Thomas and Drettakis, George},
title = {3D Gaussian Splatting for Real-Time Radiance Field Rendering},
year = {2023},
issue_date = {August 2023},
publisher = {Association for Computing Machinery},
address = {New York, NY, USA},
volume = {42},
number = {4},
issn = {0730-0301},
url = {https://doi.org/10.1145/3592433},
doi = {10.1145/3592433},
journal = {ACM Trans. Graph.},
month = jul,
articleno = {139},
numpages = {14}
}

@article{mildenhall2021nerf,
author = {Mildenhall, Ben and Srinivasan, Pratul P. and Tancik, Matthew and Barron, Jonathan T. and Ramamoorthi, Ravi and Ng, Ren},
title = {NeRF: representing scenes as neural radiance fields for view synthesis},
year = {2021},
issue_date = {January 2022},
publisher = {Association for Computing Machinery},
address = {New York, NY, USA},
volume = {65},
number = {1},
issn = {0001-0782},
url = {https://doi.org/10.1145/3503250},
doi = {10.1145/3503250},
journal = {Commun. ACM},
month = dec,
pages = {99–106},
numpages = {8}
}

@InProceedings{fridovich2022plenoxels,
    author    = {Fridovich-Keil, Sara and Yu, Alex and Tancik, Matthew and Chen, Qinhong and Recht, Benjamin and Kanazawa, Angjoo},
    title     = {Plenoxels: Radiance Fields Without Neural Networks},
    booktitle = {Proceedings of the IEEE/CVF Conference on Computer Vision and Pattern Recognition (CVPR)},
    month     = {June},
    year      = {2022},
    pages     = {5501-5510}
}

@article{muller2022instant,
author = {M\"{u}ller, Thomas and Evans, Alex and Schied, Christoph and Keller, Alexander},
title = {Instant neural graphics primitives with a multiresolution hash encoding},
year = {2022},
issue_date = {July 2022},
publisher = {Association for Computing Machinery},
address = {New York, NY, USA},
volume = {41},
number = {4},
issn = {0730-0301},
url = {https://doi.org/10.1145/3528223.3530127},
doi = {10.1145/3528223.3530127},
journal = {ACM Trans. Graph.},
month = jul,
articleno = {102},
numpages = {15}
}

@InProceedings{lassner2021pulsar,
    author    = {Lassner, Christoph and Zollhofer, Michael},
    title     = {Pulsar: Efficient Sphere-Based Neural Rendering},
    booktitle = {Proceedings of the IEEE/CVF Conference on Computer Vision and Pattern Recognition (CVPR)},
    month     = {June},
    year      = {2021},
    pages     = {1440-1449}
}

@InProceedings{guedon2024sugar,
    author    = {Gu\'edon, Antoine and Lepetit, Vincent},
    title     = {SuGaR: Surface-Aligned Gaussian Splatting for Efficient 3D Mesh Reconstruction and High-Quality Mesh Rendering},
    booktitle = {Proceedings of the IEEE/CVF Conference on Computer Vision and Pattern Recognition (CVPR)},
    month     = {June},
    year      = {2024},
    pages     = {5354-5363}
}

@InProceedings{keetha2024splatam,
    author    = {Keetha, Nikhil and Karhade, Jay and Jatavallabhula, Krishna Murthy and Yang, Gengshan and Scherer, Sebastian and Ramanan, Deva and Luiten, Jonathon},
    title     = {SplaTAM: Splat Track \& Map 3D Gaussians for Dense RGB-D SLAM},
    booktitle = {Proceedings of the IEEE/CVF Conference on Computer Vision and Pattern Recognition (CVPR)},
    month     = {June},
    year      = {2024},
    pages     = {21357-21366}
}

@InProceedings{zhang2018unreasonable,
author = {Zhang, Richard and Isola, Phillip and Efros, Alexei A. and Shechtman, Eli and Wang, Oliver},
title = {The Unreasonable Effectiveness of Deep Features as a Perceptual Metric},
booktitle = {Proceedings of the IEEE Conference on Computer Vision and Pattern Recognition (CVPR)},
month = {June},
pages={586-595},
year = {2018}
}

@misc{galindo19-NLOSDataset, 
 title={A dataset for benchmarking time-resolved non-line-of-sight imaging}, 
 author={Galindo, Miguel and Marco, Julio and O'Toole, Matthew and Wetzstein, Gordon and Gutierrez, Diego and Jarabo, Adrian},
 publisher={IEEE}, 
 booktitle={IEEE International Conference on Computational Photography (ICCP)}, 
 url={https://graphics.unizar.es/nlos}, 
 year={2019}
}

@inproceedings{malik2023transient,
 author = {Malik, Anagh and Mirdehghan, Parsa and Nousias, Sotiris and Kutulakos, Kyros and Lindell, David},
 booktitle = {Advances in Neural Information Processing Systems},
 editor = {A. Oh and T. Naumann and A. Globerson and K. Saenko and M. Hardt and S. Levine},
 pages = {71569--71581},
 publisher = {Curran Associates, Inc.},
 title = {Transient Neural Radiance Fields for Lidar View Synthesis and 3D Reconstruction},
 url = {https://proceedings.neurips.cc/paper_files/paper/2023/file/e261e92e1cfb820da930ad8c38d0aead-Paper-Conference.pdf},
 volume = {36},
 year = {2023}
}

@InProceedings{malik2025neural,
    author    = {Malik, Anagh and Attal, Benjamin and Xie, Andrew and O'Toole, Matthew and Lindell, David B.},
    title     = {Neural Inverse Rendering from Propagating Light},
    booktitle = {Proceedings of the IEEE/CVF Conference on Computer Vision and Pattern Recognition (CVPR)},
    month     = {June},
    year      = {2025},
    pages     = {10534-10544}
}

@InProceedings{behari2025blurred,
    author    = {Behari, Nikhil and Young, Aaron and Somasundaram, Siddharth and Klinghoffer, Tzofi and Dave, Akshat and Raskar, Ramesh},
    title     = {Blurred LiDAR for Sharper 3D: Robust Handheld 3D Scanning with Diffuse LiDAR and RGB},
    booktitle = {Proceedings of the IEEE/CVF Conference on Computer Vision and Pattern Recognition (CVPR)},
    month     = {June},
    year      = {2025},
    pages     = {26954-26964}
}

@InProceedings{mu2024towards,
    author    = {Mu, Fangzhou and Sifferman, Carter and Jungerman, Sacha and Li, Yiquan and Han, Mark and Gleicher, Michael and Gupta, Mohit and Li, Yin},
    title     = {Towards 3D Vision with Low-Cost Single-Photon Cameras},
    booktitle = {Proceedings of the IEEE/CVF Conference on Computer Vision and Pattern Recognition (CVPR)},
    month     = {June},
    year      = {2024},
    pages     = {5302-5311}
}

@InProceedings{sifferman2025recovering,
    author    = {Sifferman, Carter and Li, Yiquan and Li, Yiming and Mu, Fangzhou and Gleicher, Michael and Gupta, Mohit and Li, Yin},
    title     = {Recovering Parametric Scenes from Very Few Time-of-Flight Pixels},
    booktitle = {Proceedings of the IEEE/CVF International Conference on Computer Vision (ICCV)},
    month     = {October},
    year      = {2025},
    pages     = {27989-27999}
}

@ARTICLE{qu2024z,
  author={Qu, Ziyuan and Vengurlekar, Omkar and Qadri, Mohamad and Zhang, Kevin and Kaess, Michael and Metzler, Christopher and Jayasuriya, Suren and Pediredla, Adithya},
  journal={IEEE Transactions on Pattern Analysis and Machine Intelligence}, 
  title={Z-Splat: Z-Axis Gaussian Splatting for Camera-Sonar Fusion}, 
  year={2025},
  volume={47},
  number={9},
  pages={7255-7267},
  doi={10.1109/TPAMI.2024.3462290}}

@ARTICLE{sethuraman2025sonarsplat,
  author={Sethuraman, Advaith V. and Rucker, Max and Bagoren, Onur and Kung, Pou-Chun and Amutha, Nibarkavi N.B. and Skinner, Katherine A.},
  journal={IEEE Robotics and Automation Letters}, 
  title={SonarSplat: Novel View Synthesis of Imaging Sonar via Gaussian Splatting}, 
  year={2025},
  volume={10},
  number={12},
  pages={13312-13319},
  doi={10.1109/LRA.2025.3627089}}

@InProceedings{kung2025radarsplat,
    author    = {Kung, Pou-Chun and Harisha, Skanda and Vasudevan, Ram and Eid, Aline and Skinner, Katherine A.},
    title     = {RadarSplat: Radar Gaussian Splatting for High-Fidelity Data Synthesis and 3D Reconstruction of Autonomous Driving Scenes},
    booktitle = {Proceedings of the IEEE/CVF International Conference on Computer Vision (ICCV)},
    month     = {October},
    year      = {2025},
    pages     = {27596-27606}
}
